\documentstyle[12pt]{article}

\begin{document}


\def\a{\alpha}
\def\b{\beta}
\def\c{\varepsilon}
\def\d{\delta}
\def\e{\epsilon}
\def\f{\phi}
\def\g{\gamma}
\def\h{\theta}
\def\k{\kappa}
\def\l{\lambda}
\def\m{\mu}
\def\n{\nu}
\def\p{\psi}
\def\q{\partial}
\def\r{\rho}
\def\s{\sigma}
\def\t{\tau}
\def\u{\upsilon}
\def\v{\varphi}
\def\w{\omega}
\def\x{\xi}
\def\y{\eta}
\def\z{\zeta}
\def\D{\Delta}
\def\G{\Gamma}
\def\L{\Lambda}
\def\F{\Phi}
\def\P{\Psi}
\def\S{\Sigma}

\def\o{\over}

\def\IJMP{Int.~J.~Mod.~Phys. }
\def\MPL{Mod.~Phys.~Lett. }
\def\NP{Nucl.~Phys. }
\def\PL{Phys.~Lett. }
\def\PR{Phys.~Rev. }
\def\PRL{Phys.~Rev.~Lett. }
\def\PTP{Prog.~Theor.~Phys. }
\def\ZP{Z.~Phys. }


\title{
\begin{flushright}
\large UT-757
\end{flushright}
       Natural New Inflation in Broken Supergravity}

\author{Izawa K.-I. and T. Yanagida \\
        \\ {\sl Department of Physics, University of Tokyo, Tokyo
        113, Japan}}

\date{August, 1996}

\maketitle\thispagestyle{empty}
\setlength{\baselineskip}{3.6ex}

\begin{abstract}
We consider a natural new inflationary model in broken supergravity
based on an $R$ symmetry.
The model predicts a concrete relation
between the amplitude of primordial density fluctuations
and the scale of supersymmetry breaking.
The observed value of the density fluctuations is obtained
for the gravitino mass of order the weak scale
along with a power-law spectral index considerably less than one,
which may be tested in future observations.
\end{abstract}

\newpage

\section{Introduction}

Low-energy supersymmetry has attracted much attention
in particle physics, since it provides a conceivable solution to the hierarchy
problem \cite{A, Nil}.
Supersymmetric theories naturally accommodate gravity
in the form of supergravity \cite{Nil},
which may give us a consistent description
of physics below the Planck scale. However, supergravity generically allows
a constant term in a superpotential. Thus we expect a negative cosmological
constant of order the Planck scale,
which yields an anti de Sitter universe.
This leads us to seek a further symmetry
which avoids such a disastrous situation.

Supersymmetric field theories admit a peculiar symmetry called
$R$ symmetry \cite{Nil}. It is unique in that it can forbid a constant term
in a superpotential and thus restrict a cosmological constant in supergravity.
It is also ubiquitous in phenomenological models with supersymmetry.
Indeed it is a generic ingredient
for causing dynamical supersymmetry breaking \cite{Nel}.
These considerations lead us to impose an $R$ symmetry
in the framework of supergravity.

In this paper, we consider an $R$-invariant model of an inflaton
where spontaneous breakdown of the $R$ symmetry
naturally generates an inflationary universe.
The vanishing cosmological constant
in the present vacuum implies that the contributions of inflaton potential
and supersymmetry breaking sector to the vacuum energy
cancel out between
each other provided the other contributions are negligible\rlap.
\footnote{This cancellation requires a fine tuning of parameters,
which we postulate in this paper.}
Then we obtain a concrete relation between the amplitude
of primordial density fluctuations
and the scale of supersymmetry breaking in our universe.
We see that the value of primordial density fluctuations
predicted for the gravitino mass of order the weak scale is just around
that obtained in the observational analyses.
The prediction of the spectral index tends to be considerably less than one
for the observed value of the density fluctuations,
which may be tested in future observations.

\section{The model}

Let us introduce an inflaton superfield $\f$ with $R$ charge $2/(n+1)$,
where $n$ denotes a positive integer of order one.
Namely, it transforms as
\begin{equation}
 \f(\h) \rightarrow e^{i{2 \o n+1}\a} \f(e^{-i\a}\h).
\end{equation}
This charge assignment allows a tree-level superpotential
\begin{equation}
 W_0 = - {g \o n+1} \f^{n+1},
\end{equation}
where $g$ is a coupling constant of order one.
Here and henceforth, we set the gravitational scale
$M \simeq 2.4 \times 10^{18}$ GeV equal to unity
and regard it as a plausible cutoff in supergravity.
Note that the superpotential $W_0$ by itself yields a fairly flat potential
for $n \geq 3$,
which is desirable for a slow-roll inflationary scenario.
We further assume the presence of a (composite) superfield
\footnote{This $R$ charge is chosen so that the inflaton $\f$
couples to this (composite) superfield.}
with $R$ charge $2 - 2/(n+1)$
which condenses to give a tiny scale $v^2 \ll 1$.
This condensation breaks the $U(1)_R$ symmetry
\footnote{This U$(1)_R$ symmetry may be anomalous
due to a dynamical origin of the scale $v^2$, which avoids the presence
of an $R$ axion. We also note that one may impose a discrete $R$ symmetry
from the start instead of the continuous one.}
down to a discrete $R$ symmetry $Z_{2n}$ \cite{Kum}.
Then we expect an effective superpotential
\begin{equation}
 W = v^2 \f - {g \o n+1} \f^{n+1}.
 \label{ADDEQ}
\end{equation}
The $R$-invariant effective K\"ahler potential is given by
\begin{equation}
 K = |\f|^2 + {k \o 4}|\f|^4 + \cdots,
\end{equation}
where $k$ is naturally of order one and assumed to be positive.
The ellipsis denotes higher-order terms, which we may ignore
in the following analysis.

The effective potential for the field $\f$
in supergravity is given by \cite{Nil}
\begin{equation}
  V = e^K
      \left\{ \left( {\q^2 K \o \q \f \q \f^*} \right)^{-1} |DW|^2
      - 3|W|^2 \right\},
 \label{EPOT}
\end{equation}
where we have defined
\begin{equation}
 DW = {\q W \o \q \f} + {\q K \o \q \f}W.
\end{equation}
This indicates that the vacua
\footnote{Although these vacua may only correspond to local minima,
possible vacua with $|\f| \geq 1$ do not affect
the following analysis.}
with $|\f| < 1$ satisfy the condition \cite{Wei}
\begin{equation}
 DW \simeq v^2 - g \f^n = 0,
\end{equation}
which yields a vacuum
\begin{equation}
 \langle \f \rangle \simeq \left({v^2 \o g}\right)^{1 \o n}.
\end{equation}
The potential at the vacuum is given by
\begin{equation}
 \langle V \rangle = -3 e^{\langle K \rangle} |\langle W \rangle|^2
                   \simeq -3\left({n \o n+1}\right)^2v^4|\langle \f \rangle|^2,
 \label{VAC}
\end{equation}
whose magnitude is much smaller than the inflation scale $V(0) = v^4$.

We propose a scenario that the negative vacuum energy Eq.(\ref{VAC})
is canceled out by a supersymmetry-breaking effect
which gives a positive contribution $\L^4$ to the cosmological constant:
\begin{equation}
 -3\left({n \o n+1}\right)^2v^4\left|{v^2 \o g}\right|^{2 \o n} + \L^4 = 0.
 \label{VCC}
\end{equation}
This cancellation results in our flat vacuum\rlap.
\footnote{This is none other than
a fine tuning of the cosmological constant,
which is the unique unnatural point in the present model.
We do not specify the supersymmetry breaking sector
since its details are unnecessary for our purposes in this paper
(see the final section).}
We note that $\L^2 \ll v^2$ for $v^2 \ll 1$.
In the hidden sector models of supersymmetry breaking,
the scale $\L$ is chosen
so as to give a mass of the weak scale to the gravitino:
\begin{equation}
 m_{3/2} \simeq {\L^2 \o \sqrt{3}} \simeq 10^{-16}-10^{-15}.
 \label{GRA}
\end{equation}

The inflaton mass $m_\f$ in the vacuum is given by
\begin{equation}
 m_\f \simeq n|g|^{1 \o n}v^{2-{2 \o n}}.
 \label{IM}
\end{equation}
The inflaton $\f$ may have the following $R$-invariant interactions with
the ordinary light fields $\p_i$ in the K\"ahler potential:
\begin{equation}
 K(\f, \p_i) = \sum_i \l_i |\f|^2 |\p_i|^2 + \cdots,
 \label{DI}
\end{equation}
where $\l_i$ is a coupling constant of order one.
The decay width $\G_\f$ of the inflaton is then estimated as
\begin{equation}
 \G_\f \simeq \sum_i \l_i^2 |\langle \f \rangle|^2 m_\f^3.
 \label{DW}
\end{equation}
This decay results in a reheating temperature
\begin{equation}
 T_R \simeq g_*^{-{1 \o 4}}\sqrt{\G_\f},
 \label{RT}
\end{equation}
where $g_*$ is the relativistic degrees of freedom at the temperature $T_R$.
Hence we get
\begin{equation}
 T_R \simeq n^{3 \o 2} |g|^{2 \o n+1} m_{3/2}^{3n-1 \o 2(n+1)}.
 \label{RHT}
\end{equation}

\section{Inflationary Dynamics}

Let us investigate the inflationary dynamics of the above model
by means of a slow-roll approximation \cite{Kol}.

We may set $g > 0$ and $\langle \f \rangle > 0$ without loss of generality
and describe the system approximately
in terms of the inflaton field $\v$ $(\geq 0)$
which is $\sqrt{2}$ times the real part of the field $\f$.
Then the potential for the inflaton reads
\begin{equation}
 V(\v) \simeq v^4 - {k \o 2}v^4\v^2
              - {g \o 2^{{n \o 2}-1}}v^2\v^n + {g^2 \o 2^n}\v^{2n}
 \label{POT}
\end{equation}
for $\v < \langle \v \rangle = \sqrt{2} \langle \f \rangle$.
The $k$-independent contribution of $\v^2$ term
in $e^K|DW|^2$ is exactly canceled by that in $-3|W|^2$,
as was noted in Ref.\cite{Kum}.

The slow-roll inflationary regime is determined by the condition \cite{Kol}
\begin{equation}
 \e(\v) = {1 \o 2} \left({V'(\v) \o V(\v)} \right)^2 \leq 1,
 \quad |\y(\v)| \leq 1,
 \label{COND}
\end{equation}
where
\begin{equation}
 \y(\v) = {V''(\v) \o V(\v)}.
\end{equation}
For the potential Eq.(\ref{POT}), we obtain
\begin{equation}
\begin{array}{l}
 \displaystyle
 \e(\v) \simeq {1 \o 2}
          \left({-kv^4\v - {g \o 2^{{n \o 2}-1}}nv^2\v^{n-1} \o v^4}\right)^2
        = {\v^2 \o 2}(k + {g^2 \o 2^{n-1}}nv^{-2}\v^{n-2})^2, \\
 \noalign{\vskip 2ex}
 \displaystyle
 \y(\v) \simeq {-kv^4 - {g \o 2^{{n \o 2}-1}}n(n-1)v^2\v^{n-2} \o v^4}
        = -k - {g \o 2^{{n \o 2}-1}}n(n-1)v^{-2}\v^{n-2}.
\end{array}
\end{equation}
The slow-roll condition Eq.(\ref{COND}) is satisfied
for $k \leq 1$ and $\v \leq \v_f$
where
\begin{equation}
 \v_f \simeq \sqrt{2} \left({(1-k)v^2 \o gn(n-1)}\right)^{1 \o n-2},
\end{equation}
which
provides the value of the inflaton field at the end of inflation\rlap.
\footnote{A sufficiently large expansion of the universe
is achieved under the condition that the initial amplitude of the inflaton
$\v$ is localized near the origin.
We suspect that such an initial condition is derived
from some underlying physics.
For example, let us consider that the inflation scale $v^2$
in Eq.(\ref{ADDEQ}) arises from a hypercolor quark condensation
$\l \langle Q \bar{Q} \rangle = v^2$, where the inflaton couples to the
hyperquarks as $\l Q \bar{Q} \v$. If the initial values of $Q$ and $\bar{Q}$
are large as order one, the inflaton field $\v$ is set to be near the origin.}
This value is smaller than the vacuum expectation value $\langle \v \rangle$
due to $v^2 \ll 1$,
which is consistent with the approximation Eq.(\ref{POT})
of the inflaton potential
for discussing the inflationary dynamics.
The Hubble parameter during the inflation ($0 < \v \leq \v_f$) is given by
\begin{equation}
 H \simeq \sqrt{V(0) \o 3} \simeq {v^2 \o \sqrt{3}}.
 \label{HBL}
\end{equation}

Let us turn to consideration on the horizon of the present universe.
The $e$-fold number $N$ of the present horizon is given by \cite{Kol}
\begin{equation}
 N \simeq 67 + {1 \o 3}\ln H + {1 \o 3}\ln T_R
   \simeq 67 + {1 \o 3}\ln(n^{3 \o 2} m_{3/2}^{5n-1 \o 2(n+1)}).
 \label{EFLD}
\end{equation}

Let $\v_N$ be the value of the field $\v$ when the observable universe
crossed the horizon during the inflation.
Then the $e$-fold number $N$ is also given by
\begin{equation}
 N = \int_{\v_f}^{\v_N} \! d\v \, {V(\v) \o V'(\v)}.
\end{equation}
\newline
$(i)$ For $1/n \leq k < 1$, we obtain
\begin{equation}
 N \simeq \int_{\v_f}^{\v_N} \! d\v \, {v^4 \o -kv^4\v}
   = {1 \o k}\ln({\v_f \o \v_N}).
\end{equation}
That is,
\begin{equation}
 \v_N \simeq \v_f e^{-kN}.
\end{equation}
\newline
$(ii)$ For $1/N \leq k < 1/n$, we obtain
\begin{equation}
 N \simeq \int_{\bar \v}^{\v_N} \! d\v \, {v^4 \o -kv^4\v}
 + \int_{\v_f}^{\bar \v} \! d\v \, {v^4 \o -{g \o 2^{{n \o 2}-1}}nv^2\v^{n-1}}
   = {1 \o k}\ln({{\bar \v} \o \v_N}) + {1-nk \o (n-2)k(1-k)},
\end{equation}
where $\bar \v$ is determined by
\begin{equation}
 kv^4{\bar \v} = {g \o 2^{{n \o 2}-1}}nv^2{\bar \v}^{n-1}.
\end{equation}
That is,
\begin{equation}
 \v_N \simeq {\bar \v} e^{-k{\bar N}},
\end{equation}
where
\begin{equation}
 {\bar \v} = \sqrt{2} \left({kv^2 \o gn}\right)^{1 \o n-2}, \quad
 {\bar N} = N - {1-nk \o (n-2)k(1-k)}.
\end{equation}
\newline
$(iii)$ We do not consider the region  $k < 1/N$
since the coupling $k$ seems unnaturally small for
$N$ of several decades\rlap.
\footnote{Roughly speaking,
this is the situation analyzed in Ref.\cite{Kum}.}

The value $\v_N$ should exceed the amplitude of quantum fluctuations
of the inflaton
field in the de Sitter universe \cite{Kol}:
\begin{equation}
 \D \v \simeq {H \o 2\pi} \simeq {v^2 \o 2\pi \sqrt{3}}.
\end{equation}
For $n = 3$, we obtain $N \simeq 47$ from Eq.(\ref{GRA}) and Eq.(\ref{EFLD}).
The condition
\begin{equation}
 \v_N \simeq {\sqrt{2} k v^2 \o 3g} \exp(-kN + {1-3k \o 1-k}) > \D \v
\end{equation}
implies that $k$ seems too small to be natural for $g$ of order one.
Hence we discard this possibility and restrict ourselves to $n \geq 4$,
where the condition $\v_N > \D \v$ is satisfied for a natural range
of the parameter $k$.

\section{The Density Fluctuations and Spectral Index}

In the above inflationary model,
the amplitude of primordial density fluctuations $\d \r / \r$,
which arises from quantum fluctuations $\D \v$
of the inflaton field, is given by \cite{Kol}
\begin{equation}
 {\d \r \o \r} \simeq {3 \o 5 \pi}{H^3 \o |V'(\v_N)|}
 \simeq {1 \o 5\sqrt{3}\pi}{V^{3 \o 2}(\v_N) \o |V'(\v_N)|}
\end{equation}
and the spectral index $n_s$ of the density fluctuations
is given by \cite{Kol}
\begin{equation}
 \label{X}
 \begin{array}{l}
 \displaystyle
  n_s \simeq 1 - 6\e(\v_N) + 2\y(\v_N) \\
 \noalign{\vskip 1ex}
 \displaystyle
  \ \quad
  \simeq 1 - 2k\left\{1+(n-1)\exp\left[-k(n-2)N+{1-nk \o 1-k}\right]\right\}.
 \end{array}
\end{equation}
For $1/N \ll k < 1$, we obtain $n_s \simeq 1-2k$.
The lower bound of the tilt allowed by observations
implies $n_s > 0.6 $ \cite{COBE, ST},
which is realized for $k < 0.2$.
Thus we adopt the range $1/N \leq k < 0.2$
and evaluate the density fluctuations
by means of an input Eq.(\ref{GRA}).
\newline
({\it a}) For $n = 4$, we obtain
\begin{equation}
 N \simeq 45,
\end{equation}
which gives
\begin{equation}
 8 \times 10^{-6} g^{3 \o 5} \leq 2 \times 10 g^{3 \o 5} m_{3/2}^{2 \o 5}
 \leq {\d \r \o \r} < 5 \times 10^3 g^{3 \o 5} m_{3/2}^{2 \o 5}
 \leq 5 \times 10^{-3} g^{3 \o 5}.
\end{equation}
The lower and upper bounds correspond to the cases of
$k=1/45$, $m_{3/2}=10^{-16}$ and $k=0.2$, $m_{3/2}=10^{-15}$, respectively.
\newline
({\it b}) For $n \geq 5$, we obtain
\begin{equation}
 0 < {\d \r \o \r} < 2 \times 10^3 g^{4 \o 9} m_{3/2}^{5 \o 9}
 \leq 1 \times 10^{-5} g^{4 \o 9}.
\end{equation}

The observational data yield $\d \r / \r \simeq 2 \times 10^{-5}$
\cite{COBE, ST},
which implies that a realistic inflationary model is given
\footnote{In the case of $n = 5$, the required value
$\d \r /\r \simeq 2 \times 10^{-5}$ implies that the spectral index
$n_s \simeq 0.6$, which may be marginally consistent with the observations.}
in the case of $n=4$.
Then the required amplitude of the density fluctuations is obtained
for $k \simeq 0.03 - 0.13$ and $g$ of order one,
which results in the spectral index
$n_s \simeq 0.91 - 0.74$. This tilted power spectrum
of the primordial density fluctuations
may be adequate for structure formation in our universe \cite{ST}.

\section{Conclusion}

Let us summarize the model with $n=4$.
For the gravitino mass Eq.(\ref{GRA}) of the weak scale,
we obtain the inflation scale $v \simeq 10^{-6}$
and the Hubble parameter during the inflation
$H \simeq 10^{-12}$ from Eq.(\ref{VCC}) and Eq.(\ref{HBL}).
The inflaton mass Eq.(\ref{IM}) and the reheating temperature Eq.(\ref{RHT})
turn out to be
\begin{equation}
 m_\f \simeq 10^{-9}, \quad T_R \simeq 10^{-16}.
\end{equation}
For the coupling $k \simeq 0.1$ in the K\"ahler potential,
we get the amplitude $\d \r/\r \simeq 10^{-5}$.
The model with the observed amplitude
$\d \r/\r \simeq 2 \times 10^{-5}$
predicts a tilted power spectrum
of the primordial density fluctuations with the index $n_s \simeq 0.8$.

If the mass of some right-handed neutrino is less than $m_\f/2$,
the inflaton can decay to a pair of the neutrinos
through the interaction Eq.(\ref{DI}). In that case,
baryogenesis may be followed by
leptogenesis from decay of the right-handed neutrino \cite{Fuk}
since $T_R$ is of order the weak scale in the above model.
The right-handed neutrino with the mass of this order induces,
through the seesaw mechanism \cite{Yan},
a tiny mass of a left-handed neutrino in an interesting range for the solution
to the solar neutrino problem \cite{F}.
We note that the reheating temperature is possibly higher than the weak scale
when the inflaton field is involved in a stronger interaction than
the one in Eq.(\ref{DI}) \cite{Kum}.

Let us comment on the supersymmetry breaking sector.
The $\L^4$ contribution
to the vacuum energy in Eq.(\ref{VAC}) is obtained, for example, by
introducing a superfield $Z$ and its superpotential $W(Z) = \L^2 Z$
with an origin of the scale $\L$ presumably dynamical \cite{Iza}.
During inflation $\v \simeq 0$, the field $Z$ acquires a mass
of the Hubble scale,
which keeps the condition $Z \simeq 0$
at the inflationary epoch
\footnote{This may result in a supersymmetry-breaking vacuum
without the so-called Polonyi problem \cite{Ban}
if the vacuum lies near the origin $\langle Z \rangle \simeq 0$.}
and the contribution of the $Z$ sector is negligible during the inflation
for $\L^2 \ll v^2$. Thus
the introduction of the field $Z$
scarcely affects the inflationary dynamics.
We note that the gravitino mass is possibly as light
\footnote{Such a light gravitino ($m_{3/2} \simeq 10^{-24}$)
is realized in the framework of dynamical supersymmetry breaking
at low energies. If $n_s > 0.7$ is confirmed in future observations,
we will see that
our scenario of the vacuum-energy cancelation may be incompatible with
the low-energy supersymmetry breaking.}
as $10^{-24}$ for the lower
bound of the spectral index $n_s \simeq 0.6$ though we have regarded
$m_{3/2}$ as the weak scale throughout the paper.
On the other hand, if we consider the case that
the contribution of the inflaton
potential to the cosmological constant is canceled by some GUT scale physics
instead of the supersymmetry breaking sector, the vacuum expectation value
of the inflaton $\langle \v \rangle$ turns out to be of order one.
Then the present model realizes supersymmetric topological inflation:
The model possesses
the discrete $R$ symmetry $Z_{2n}$,
which is spontaneously broken to the $R$ parity symmetry
by the inflaton condensation $\langle \v \rangle \neq 0$.
Thus we have $n$ degenerate vacua in this model, which cause domain
wall structures in the whole universe.
The initial value of the inflaton field $\v \simeq 0$ may
be naturally achieved
in a sufficiently large region inside a domain wall
for topological reasons and the resultant
defects serve as seeds for inflation \cite{Lin}.

\section*{Acknowledgement}

We would like to thank E.D.~Stewart for valuable comments.

\end{document}